\documentstyle{prhep97}
\begin{document}
\begin{titlepage}
\begin{centering}

\vspace{5cm}

{\Large\bf Low Energy Tests of the Standard Model}\\

\vspace{15pt}

{\Large\bf from $\beta$-Decay and Muon Capture}\footnote{Contribution
to the Proceedings of {\sl The International Europhysics Conference
on High-Energy Physics\/}, 19--26 August 1997, Jerusalem, Israel}

\vspace{2cm}

Jan Govaerts\\ 

\vspace{0.5cm}

{\em Institute of Nuclear Physics}
\\
{\em Catholic University of Louvain}
\\
{\em 2, Chemin du Cyclotron}
\\
{\em B-1348 Louvain-la-Neuve, Belgium}\\

\vspace{30pt}

\begin{center}
{\bf Abstract}
\vspace{20pt}
\begin{quote}
Two recent low energy precision experiments are considered, in order
to illustrate how limits set by these measurements for couplings
beyond the Standard Model are complementary to high energy constraints.
\end{quote}
\end{center}
\end{centering} 

\vspace{5cm}

\noindent UCL-IPN-97-P03\\
\noindent hep-ph/9711496\\
November 1997

\end{titlepage}

\twocolumn 

\makeatletter
\let\chapter\hid@chapter
\makeatother


\authorrunning{J. Govaerts}
\titlerunning{{\talknumber}: SM Low Energy Tests}
 

\def\talknumber{702} 

\title{{\talknumber}: Low Energy Tests of the Standard Model from
$\beta$-Decay and Muon Capture }
\author{Jan Govaerts\ 
(govaerts@fynu.ucl.ac.be)}
\institute{Institute of Nuclear Physics\\
Catholic University of Louvain\\
2, Chemin du Cyclotron\\
B-1348 Louvain-la-Neuve\\
Belgium}

\maketitle

\begin{abstract}
Two recent low energy precision experiments are considered,
in order to illustrate how limits set by these measurements
for couplings beyond the Standard Model are com\-ple\-men\-ta\-ry to
high energy constraints.
\end{abstract}
\section{Introduction}

Low energy precision experiments are complementary to high energy ones, both
in the diversity of available experimental techniques as well as in the
probed range of the parameter spaces of different model extensions of the
Standard Model (SM).

It is the purpose of the present contribution to highlight such
complementarity with the example of two recent precision measurements,
one in nuclear $\beta$-decay, and the other in nuclear muon capture
on $^3$He. The latter measurement constrains dif\-fe\-rent hadronic
corrections to the char\-ged electroweak current within the SM,
while both measurements constrain a variety of
model extensions of the SM. Some of these constraints
appear to be new, or to be com\-pe\-ti\-ti\-ve with existing ones derived
from high energy measurements \cite{Proc}.

\section{Nuclear $\beta$-decay}

The polarisation-asymmetry correlation in $\beta$-decay presents an
increased sensitivity to any deviation from the $(V-A)$
structure of the char\-ged electroweak interaction \cite{Quin1}.
Such experiments consist in the measurement of the longitudinal
polarisation of the $\beta$ particle emitted in a direction either
antiparallel or parallel to the polarisation of an oriented nucleus.
The ratio of these two po\-la\-ri\-sa\-tions---a re\-la\-ti\-ve measurement 
less prone to systematic cor\-rec\-tions---is of the form \cite{Quin1,Gov1}
$R(J)=R_0(J)\,\left[\,1-k(J)\,\Delta\,\right]$,
$J$ being the degree of nuclear polarisation, while $R_0(J)$ and $k(J)$
are known functions of the $\beta$ energy and asymmetry, quantities which
are experimentally accessible. Finally, $\Delta$ is a vanishing quantity in the
SM, which may be expressed in terms of the underlying effective
charged current interaction \cite{Gov1}. 
In particular, for $J$ close to unity, the factor
$k(J)$ can become appreciable, thus enhancing the sensitivity
to a possible deviation $\Delta\ne 0$.

Two such experiments have been performed, one using polarised
$^{12}$N \cite{Allet,Thomas}, the other polarised $^{107}$In \cite{Seve}.
The combined data result in the precision value \cite{Thomas},
\begin{equation}
\Delta=0.0004\pm 0.0026\ \ \ ,
\label{eq:Delta}
\end{equation}
thus in perfect agreement with the SM prediction. Prospects are to
perform a similar measurement for $^{17}$F at ISOLDE/CERN, as well as
for $\mu^+$ decay at PSI, with at least a 50-fold improvement in the
precision of the Michel parameter $\xi''$.

\section{Nuclear muon capture}

The statistical muon capture rate on $^3$He to the triton channel
was measured in a recent experiment to a precision of 0.3\%.
The experimental result is \cite{He3},
\begin{equation}
\lambda_{\rm exp}=1496\pm\,4\ {\rm s}^{-1}\ \ \ ,
\label{eq:rate}
\end{equation}
to be compared to the
prediction \cite{Jim1} $\lambda_{\rm theor}=1497\pm 12$ s$^{-1}$.
Prospects are to perform at PSI a similar measurement for muon capture
on the proton to better than 1\%, with as by-product a 2-fold improvement
in the precision of Fermis' coupling constant $G_F$.

\section{Tests within the SM}

The result (\ref{eq:rate}) allows for tests of QCD chiral symmetry
predictions, namely tests of PCAC and of second-class
currents\footnote{For further details, see Ref.\cite{Gov2}.}.
The vector and axial current matrix elements are
pa\-ra\-me\-tri\-sed in terms of six nuclear form factors. These include the
pseudoscalar one $F_P$ whose value is related to that of the axial one
$F_A$ through PCAC, as well as the second-class ones, na\-me\-ly the scalar
and tensor form factors $F_S$ and $F_T$, which vanish in the limit
of exact isospin and charge conjugation invariance. Using CVC and
the $^3$H $\beta$-decay rate, the values of the remaining form factors
may be determined with sufficient reliability \cite{Jim1}.

On that basis, ignoring first the contributions of $F_S$ and $F_T$,
the result (\ref{eq:rate}) leads to a value
for $F_P$ which, when compared to the PCAC prediction, is in a ratio
of \cite{He3} $1.004\pm 0.076$. At the level of the nucleon, the same 
ratio is then \cite{Jim2} $1.05\pm 0.19$. Consequently,
the value (\ref{eq:rate}) provides the most precise test of nuclear PCAC
available, a situation
to be contrasted with the recent result \cite{Jonk}
from radiative muon capture on the proton which deviates by more than a factor
of 1.5 from the PCAC prediction.

On the other hand, assuming the PCAC value for $F_P$, (\ref{eq:rate})
may be used to set a value either for $F_S$ or for $F_T$, ignoring
in each case the contribution of the other second-class form factor.
One then obtains\footnote{The normalisation of these form factors is
relative to $q^\mu/(2M)$, $M$ being the average $^3$He-$^3$H mass.}
\cite{Gov2} $F_S=-0.062\pm 1.18$ or $F_T=0.075\pm 1.43$,
values which agree of course with expectations \cite{Shio},
and do improve on the existing situation \cite{Hols1}.

\section{Tests beyond the SM}

Physics beyond the SM may be pa\-ra\-me\-tri\-sed in terms of 4-fermion 
effective interactions at the quark-lepton level. For muon decay,
a representation in the charge exchange form has become standard
in terms of effective couplings $g^{S,V,T}_{\eta_1\eta_2}$, where the
lower indices indicate the chiralities of the electron and muon,
respectively \cite{PDG}. Similarly, $\beta$-decay is pa\-ra\-me\-tri\-sed
in terms of coupling coefficients $f^{S,V,T}_{\eta_1\eta_2}$,
with the index $\eta_2$ being the chirality of the down quark,
while muon capture is pa\-ra\-me\-tri\-sed in terms of coefficients
$h^{S,V,T}_{\eta_1\eta_2}$, $\eta_1$ (resp. $\eta_2$) being the
muon (resp. $d$ quark) chirality \cite{Gov2}.

Assuming only vector and axial couplings $f^V_{\eta_1\eta_2}$, 
the result (\ref{eq:Delta}) implies
$|f^V_{RR}-f^V_{RL}|^2=0.0004\pm 0.0026$.
The quantity $\Delta$ also involves scalar and tensor
contributions, but the ensuing limits do not improve existing 
constraints \cite{Thomas}.

Under different assumptions, the result (\ref{eq:rate}) sets
new constraints \cite{Gov2}. 
L-han\-ded vector couplings only imply
$|h^V_{LL}/f^V_{LL}|^2=0.9996\pm 0.0083$,
na\-me\-ly a universality test which at present does not improve
the usual such test from pion decay\footnote{Precise to better than 0.4\%. }.
For both L- and R-handed vector couplings, one finds
$h^V_{LR}=0.0005\pm 0.0102$.
Scalar (resp. pseudoscalar) couplings are such that
$(h^S_{RR}+h^S_{RL})G_S=-0.0012\pm 0.022$,
(resp. $(h^S_{RR}-h^S_{RL})G_P=-0.078\pm 1.49$),
$G_S$, $G_P$ being the nuclear matrix elements for
the scalar and pseudoscalar quark densities.
And for tensor couplings, one has
$h^T_{RL}\,G_T/2=-0.00008\pm 0.00143$
($G_T$ being the tensor nuclear matrix element).
The scalar constraint is quite stringent, but the tensor one
is especially restrictive.

Within specific model extensions of the SM, these results translate
into constraints on the parameters of such models. The involved
observables being different from those accessible u\-su\-al\-ly from high
energy experiments, the probed regions of these parameter spaces
are complementary to one another. Here, only a few such instances
are indicated \cite{Thomas,Gov2}.

Within left-right symmetric mo\-dels not manifestly
symmetric bet\-ween their two chiral sectors, as a function of the
heavier char\-ged gauge boson mass $M_2$, the result (\ref{eq:Delta})
probes regions in the right-handed mixing matrix element $V^R_{ud}$
or in the ratio $g_R/g_L$ of gauge couplings constants which are
inaccessible \cite{Thomas,Gov3} to the collider experiments \cite{FERM}.
In particular, the latter are totally insensitive to a mass $M_2$ close 
to the $W$ mass provided for example the $V^R_{ud}$ quark mixing 
is sufficiently small, while the ratio $V^R_{ud}/V^L_{ud}$ is much
constrained in that mass region by the result (\ref{eq:Delta}) \cite{Thomas}. 
Such a possibility
is thus still worth exploring also at high energies.
The result (\ref{eq:rate}) only con\-strains the charged
gauge boson mixing angle, to a level comparable to
existing limits \cite{Gov2}.

Contact interactions are analysed similarly, replacing
the coupling coefficients
by $\pm 4\pi/\Lambda^2_{\eta_1\eta_2}$.
The result (\ref{eq:Delta}) translates into
$\Lambda^V_{R\eta_2}> 2.5$ TeV (90\% CL) for charged vector interactions
within the first generation. Note that these limits as such are not 
directly accessible to unpolarised high energy mea\-su\-re\-ments.
Eq.(\ref{eq:rate}) and $h^V_{LR}$ imply
$\Lambda^V_{LR}> 4.9$ TeV (90\% CL) for charged vector interactions
between the first quark ge\-ne\-ra\-tion and the second lepton ge\-ne\-ra\-tion.
Eq.(\ref{eq:rate}) also sets li\-mits for such scales associated
to scalar or tensor interactions. For $G_T=1$,
one has $\Lambda^T_{RL}> 9.3$ TeV (90\% CL).
These limits on contact interactions for charged currents interactions
wi\-thin the first or the first two ge\-ne\-ra\-tions are certainly comparable
to existing ones, if not better or altogether new in some cases.
Most analyses of contact interactions based on the
excess of large $Q^2$ events at HERA have concentrated on neutral current
interactions, for which the limits are in the $2.5-3.0$ TeV range \cite{Proc}.

When extending the by-now standard approach of Ref.\cite{Buch} 
for leptoquarks with a right-handed neutrino
for each generation, a new scalar and a new vector leptoquark is
possible, with three new coupling coef\-ficients for each.
The result (\ref{eq:Delta})
sets a limit on the coupling of the $S_0$ or $V_0$ leptoquarks
(in the notation of Ref.\cite{Davi}) to $(\nu_e)_R$,
namely $|\lambda^R_{S_0}\lambda^{\nu_R}_{S_0}/M^2(S_0)|$ and
$2|\lambda^R_{V_0}\lambda^{\nu_R}_{V_0}/M^2(V_0)|$ each less than
4.1 TeV$^{-2}$ (90\% CL),
limits which obviously are not available so far from high energy
measurements. The result (\ref{eq:rate}) sets
stringent limits on couplings and masses for interactions
between the first two generations, some of which improve existing
limits \cite{Davi}. This is especially
true for the ef\-fec\-ti\-ve tensor interactions induced by the 
$S_0$ and $S_{1/2}(Q=-2/3)$ lep\-to\-quarks, leading to
$|\lambda^L_{S_0}\lambda^R_{S_0}/M^2(S_0)|\,|G_T|$ and\\
$|\lambda^L_{S_{1/2}}\lambda^R_{S_{1/2}}/M^2(S_{1/2})|\,|G_T|$ 
both being less than 0.29 TeV$^{-2}$ (90\% CL).

\end{document}